\documentclass[twocolumn]{aa}
\usepackage{graphicx}
\usepackage{amsmath}
\usepackage{subfigure}
\usepackage{cite}
\usepackage{txfonts}
\usepackage{myown}   
\begin{document}
   \title{High-resolution study of a star-forming cluster \\ in the Cep-A HW2
     region}

   \subtitle{}

   \author{C.~Comito
          \inst{1}
          \and
          P.~Schilke
          \inst{1}
          \and
          U.~Endesfelder
          \inst{1}
          \and
          I.~Jim{\' e}nez-Serra
          \inst{2}
          \and
          J.~Mart{\'{\i}}n-Pintado
          \inst{2}
          }

   \offprints{C. Comito}
   
   \institute{Max-Planck-Institut f\"ur Radioastronomie, Auf dem
     H\"ugel 69,
     D-53121 Bonn, Germany\\
     \email{ccomito@mpifr-bonn.mpg.de} 
     \and Instituto de Estructura de
     la Materia, Consejo Superior de Investigaciones Cient{\'{\i}}ficas,
     Departamento de Astrof{\'{\i}}sica Molecular e Infrarroja, C/Serrano
     121, E-28006 Madrid, Spain}

   \date{Received; accepted}

 
  \abstract {Due to its relatively small distance (725~pc), the Cepheus A East
  star-forming region is an ideal laboratory to study massive star formation
  processes.}  {Based on its morphology, it has been suggested that the
  flattened molecular gas distribution around the YSO HW2 may be a
  350-AU-radius massive protostellar disk. Goal of our work is to ascertain
  the nature of this structure.}  {We have employed the Plateau de Bure
  Interferometer\footnote{Based on observations carried out with the IRAM
  Plateau de Bure Interferometer. IRAM is supported by INSU/CNRS (France), MPG
  (Germany) and IGN (Spain).} to acquire (sub-)arcsecond-resolution imaging of
  high-density and shock tracers, such as methyl cyanide (\CHTHCN) and silicon
  monoxide (SiO), towards the HW2 position.}  {On the $1\arcsec$ ($\sim
  725$~AU) scale, the flattened distribution of molecular gas around HW2
  appears to be due to the projected superposition, on the plane of the sky,
  of at least three protostellar objects, of which at least one is powering a
  molecular outflow at a small angle with respect to the line of sight. The
  presence of a protostellar disk around HW2 is not ruled out, but such
  structure is likely to be detected on a smaller spatial scale, or using
  different molecular tracers.}{}


   \maketitle
%

\section{Introduction}\label{sec:intro}

Several theories are being considered to explain the formation of massive (M
  $\geq 8$ M$_\odot$) stars, which can be roughly grouped into
  accretion-driven and coalescence-driven models (cf. Stahler et
  al. \citen{stahler2000}). In the latter case, high-mass stars would form by
  merging of two or more lower-mass objects, making the presence of stable
  massive accretion disks around the protostar very unlikely. However, only
  models based on disk-protostar interactions are capable of explaining the
  existence of jets and outflows: hence, the high incidence, in large samples
  of massive YSOs, of highly collimated outflows (cf.  Beuther et al.
  \citen{beuther2002b}) has been interpreted as indirect evidence for the
  existence of high-mass disks.
  
  It is undoubted that the direct detection of accretion onto massive
  protostars through rotating disks constitutes an important tile in the
  massive-star-formation-theory mosaic. From an observational point of view,
  this task is made very difficult by two factors: {\it i)} massive
  star-forming regions typically are far away, a few kpc on average, making
  the direct observation of small-scale structure such as disks virtually
  impossible with current instruments; and {\it ii)}, massive stars form in
  clusters, making the surrounding region extremely complex, both spatially
  and kinematically.


Located only $\sim 725$~pc from the Sun (Johnson \citen{johnson1957}),
  Cepheus~A is considered a very promising candidate for the detection of a
  massive disk. Its well-studied bipolar outflow (cf.  G\'omez et al.
  \citen{gomez1999}, hereafter G99, and references therein) is thought to be
  powered by the radio-continuum source HW2 ($\sim 10^4$ L$_{\odot}$,
  Rodr\'{\i}guez et al.  \citen{rodriguez1994}).  Curiel et
  al. (\citen{curiel2006}) report the presence of very large tangential
  velocities in the HW2 radio jet, consistent with HW2 being a massive Young
  Stellar Object (YSO).  The distribution of \WAT\ masers (Torrelles et
  al. \citen{torrelles1996}) and of the SiO emission (G99) around HW2, both
  oriented perpendicularly with respect to the direction of the flow, have
  been interpreted as strongly supporting the existence of accretion shocks
  onto a rotating and contracting molecular disk of $\sim 700$-AU diameter,
  centered on HW2, with the northeast-southwest outflow being triggered by the
  interaction between such disk and HW2 itself. Similar conclusions have been
  reached by Patel et al. (\citen{patel2005}), based on SMA observations of
  \CHTHCN\ and dust emission. However, the fact that the HW2 vicinities are
  crowded with YSOs (at least three within an area of $0\farcs6 \times
  0\farcs6$, Curiel et al. \citen{curiel2002}), together with the recent
  detection of an internally heated hot core within $0\farcs4$ from the center
  of the outflow (Mart{\'{\i}}n-Pintado et al. \citen{martinpintado2005},
  hereafter MP05) cast some doubts on this interpretation.
Based on our PdBI observations, we conclude that, on the $1\arcsec$ scale, the
  elongated molecular structure around HW2 can be explained with the
  superposition, on the plane of the sky, of at least three different
  hot-core-type sources, at least one of them being the exciting source for a
  second molecular outflow.

\section{Observations}\label{sec:obs}

In 2003 and 2004, with the Plateau de Bure Interferometer, we have carried out
observations of several high-density and shock tracers (also cf. Schilke et
al., in prep.), among which silicon monoxide (SiO) and methyl cyanide
(CH$_3$CN), towards the HW2 position ($\alpha_{\rm J2000} =22^{\rm h}56^{\rm
m}17.9^{\rm s}$, $\delta_{\rm J2000} = +62^{\circ}01\arcmin49.6\arcsec$). A
combination of high-spectral-resolution correlator units were employed to
achieve a channel width $\Delta {\rm v}$ of up to $\sim 0.3$~\kms. The five
antennas in AB (extended) configuration provided a HPBW of $2\arcsec \times
1\farcs6$ for SiO(2-1) at 86 GHz, and of $0\farcs9 \times 0\farcs7$ for
\CHTHCN($12-11$) at 220 GHz.  The data cubes were produced with natural
weighting. All maps have been CLEANed.  Analysis of all molecular spectra has
been performed after subtraction of the continuum emission.



\begin{figure*}[htbp]
\centering \subfigure{ \includegraphics[bb= 50 250 565
745,clip,angle=-90,width=6cm]{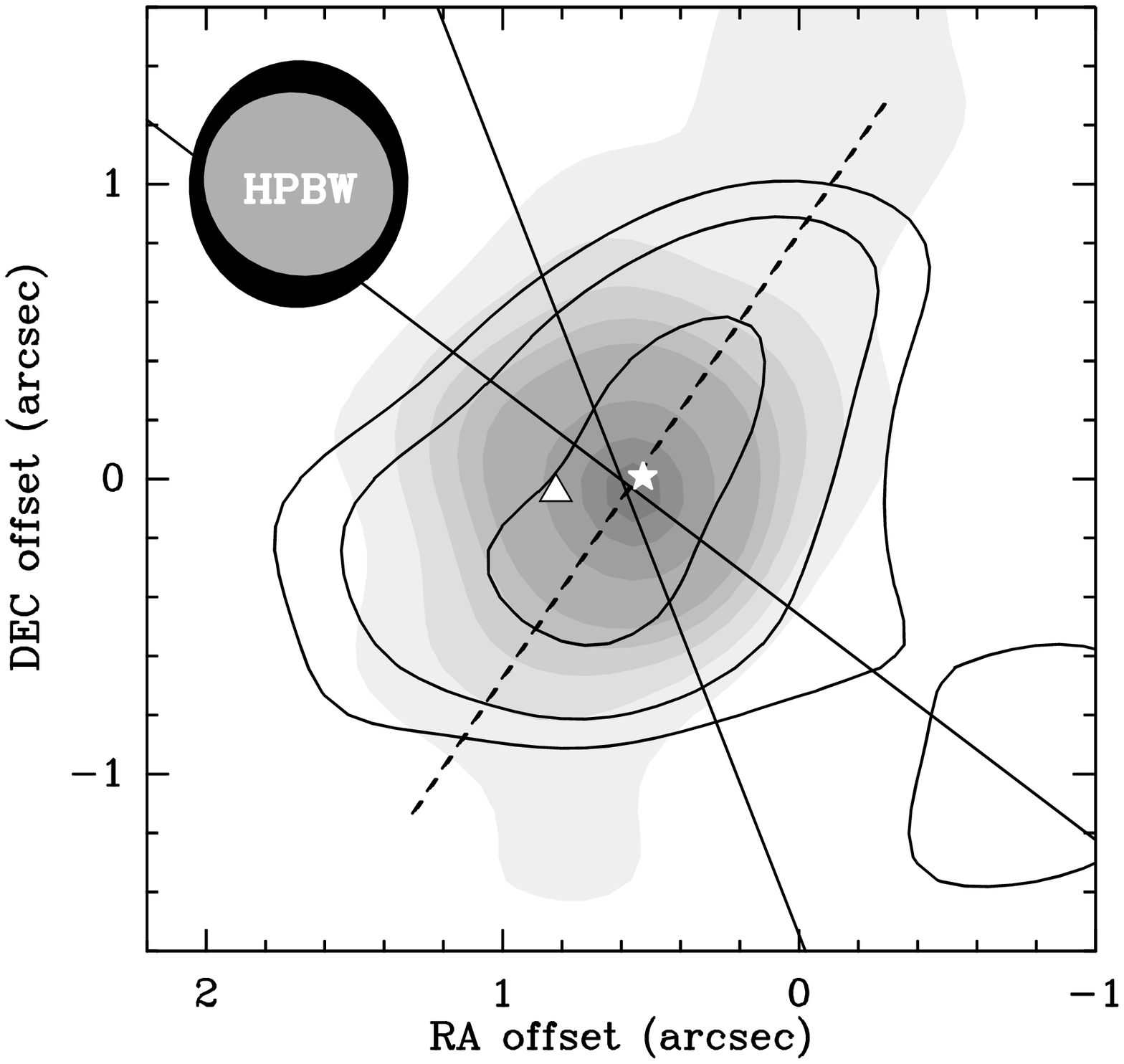}}
\hspace{-0.5cm}
\subfigure{
\includegraphics[bb= 50 250 565 745,clip,angle=-90,width=6cm]{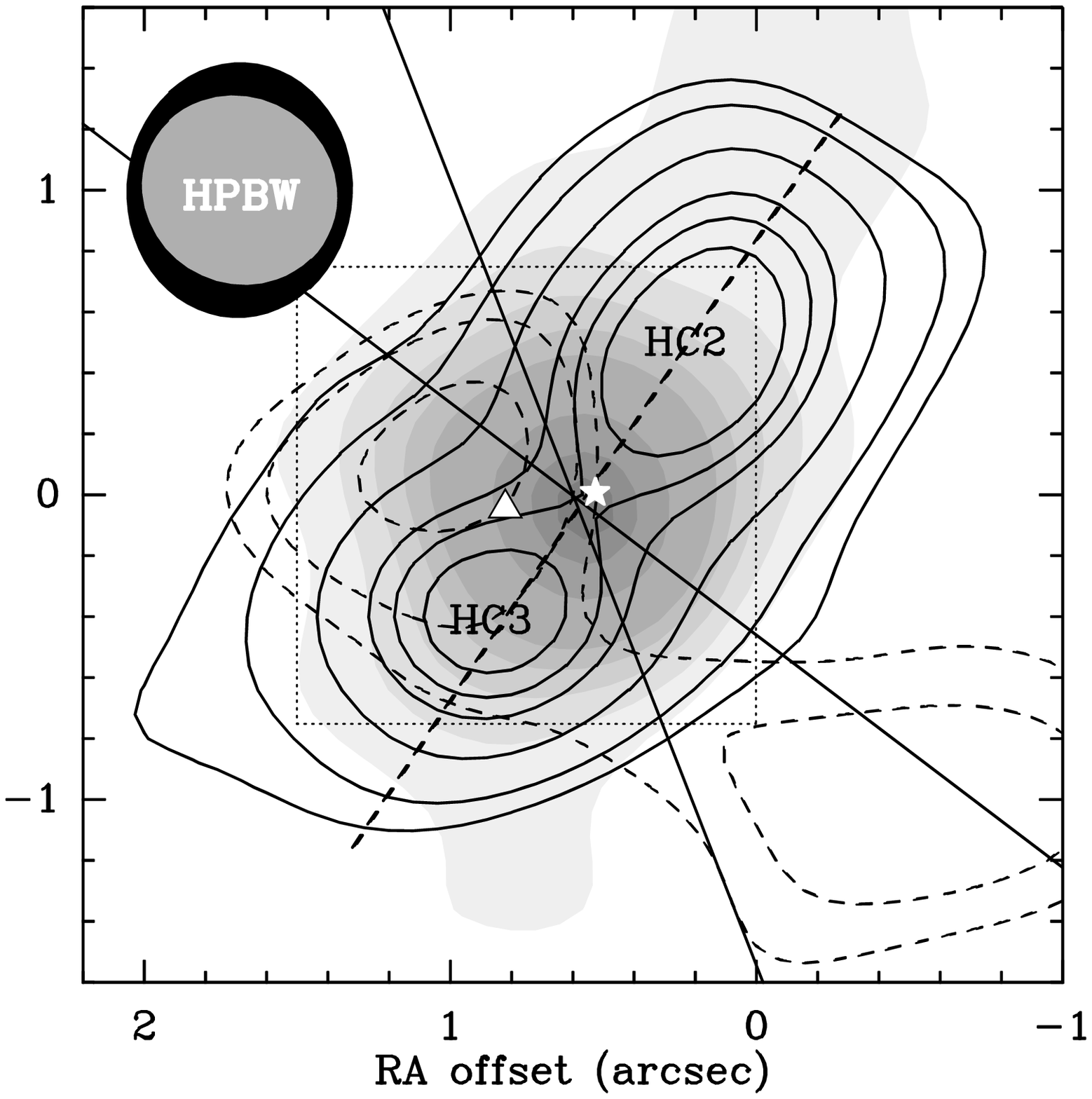}}
\hspace{-0.5cm}
\subfigure{
\includegraphics[bb= 50 250 565 745,clip,angle=-90,width=6cm]{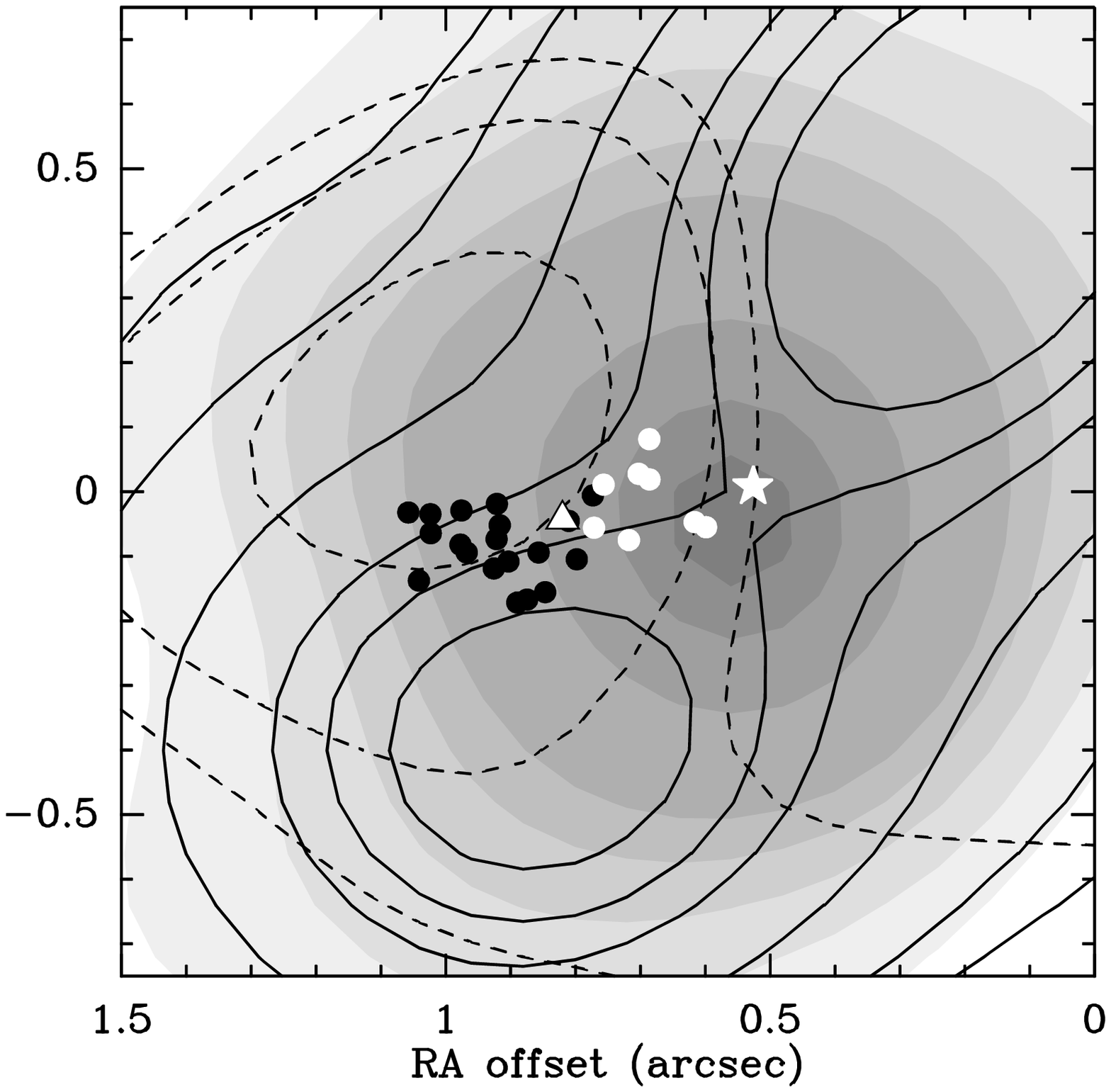}}
\caption{{\it All panels:} the grey levels represent the continuum emission at
  241~GHz. Lowest level is $3.3$~mJy/beam or $2\sigma$, highest is
  $22\sigma$. The HW2 position is indicated with a white star. The solid,
  crossing lines show the opening angle of the large-scale northeast-southwest
  outflow, inferred from HCN and \THCO\ observations obtained with PdBI
  together with the core-tracing data. The contours trace the
  continuum-subtracted \CHTHCN\ emission at 220~GHz, and in the top left
  corner, the HPBW for the continuum (grey foreground) and \CHTHCN\ (black
  background ellipse) are shown. The position of the intermediate-mass
  protostar HC (MP05) is indicated with a white triangle. {\it Left panel:}
  the integrated emission of the \CHTHCN(12$_3$-11$_3$) is shown in solid
  contours. {\it Center panel:} integrated intensity of \CHTHCN(12$_3$-11$_3$)
  between $-7$ and $-3$~\kms\ (solid contours) and between $-11.5$ and
  $-7.5$~\kms\ (dashed contours). The dotted square box indicates the area
  enlarged in the right panel. {\it Right panel:} The circles show the
  centroid positions of the SiO(2-1) emission for every channel in the range
  $-25 <$~\vlsr\ $-3.5$~\kms, with a channel width of $0.5$~\kms. The black
  circles stand for the blue-shifted (\vlsr~$< -10$~\kms), white circles for
  the red-shifted (\vlsr~$> -10$~\kms) emission. }\label{fig:ch3cn_nosio}
\end{figure*}

\section{Results}\label{sec:results}

Fig.~\ref{fig:ch3cn_nosio} (left panel) shows the Cep-A star-forming region
within a $1100$-AU radius from HW2. The peak of the 241-GHz dust emission
(grey scale) roughly coincides with the HW2 position and with the center of
the large-scale outflow. The integrated \CHTHCN\ emission is also centered on
HW2 (contours), and somewhat elongated almost perpendicularly to the direction
of the large-scale outflow. Like other molecular tracers (cf. Brogan et
al. \citen{brogan2007}), \CHTHCN\ displays two different velocity components,
centered around $-5$ and $-10$~\kms\ respectively. The solid contours in
Fig.~\ref{fig:ch3cn_nosio}, center panel, show the emission of the
\CHTHCN($12_3-11_3$) transition, integrated between $-7$ and $-3$~\kms,
whereas the emission in the range between $-11.5$ and $-7.5$~\kms\ is
represented by the dashed contours (see \S~\ref{sec:ch3cn}).

The center of SiO emission, instead, is at $\sim -10$~\kms. Silicon monoxide
peaks about $0\farcs4$ eastwards of HW2, at a position that coincides with the
HC source of MP05 (triangle in Fig.~\ref{fig:ch3cn_nosio}, see
\S~\ref{sec:sio}), close to the $-10$-\kms\ \CHTHCN\ component. In what
follows, we will discuss in more detail the SiO and \CHTHCN\ data.

\subsection{SiO}\label{sec:sio}

\begin{figure}[htbp]
\centering
\includegraphics[bb= 170 16 575 812,clip,angle=-90,width=9cm]{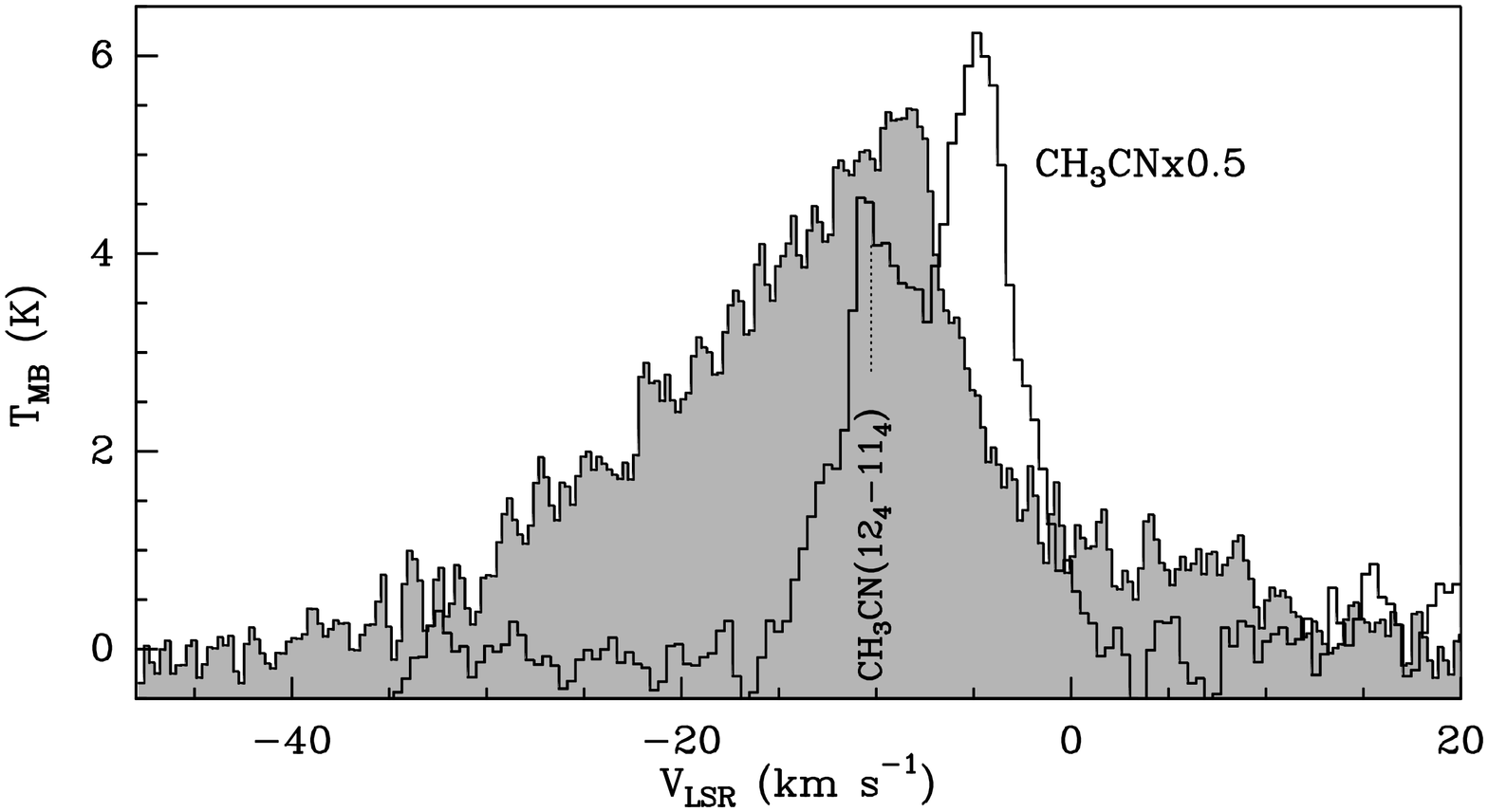}
\caption{ Comparison between the line profiles of SiO(2-1) (grey spectrum) and
\CHTHCN($12_4-11_4$) (transparent spectrum). At \vlsr$=\sim -10$~\kms, the
center of the SiO emission coincides with the position of source HC of MP05
(cf.~\S~\ref{sec:ch3cn} and Fig.~\ref{fig:ch3cn_nosio}). The SiO(2-1) spectrum
arises from a region of about $1\arcsec$ radius around the HC position
(50\%-of-peak-emission level). The \CHTHCN\ emission arises from the HC3
position (cf. ~\S~\ref{sec:ch3cn} and Fig.~\ref{fig:ch3cn_spectra}), and has
been scaled by a factor $0.5$ for a better visual comparison. No recognizable
counterpart to the $\sim -5$-\kms\ \CHTHCN\ component is observed for
SiO.}\label{fig:sio_spectrum}
\end{figure}

Our dataset confirms that the spatial distribution of this shock tracer is
mainly concentrated in the HW2 region (its presence in the large-scale outflow
is limited to a few bullets at large distances from the center), although not
centered on the HW2 position. This does indeed suggest that shock processes
are taking place in the (projected) immediate vicinities of HW2. However, if
the SiO emission were arising from accretion shocks onto a rotating disk (as
proposed by G99), we would expect to observe a similar velocity structure to
that observed for the other molecular tracers peaking around HW2. Instead, SiO
seems to be tracing a completely different kinematic component: unlike any
other line in our dataset, the SiO(2-1) line has a velocity spread of at least
$35$~\kms at the zero-flux level ($\sim 15$~\kms\ FWHM). A mass of about
90~\Msun\ would be required to produce such large line width in a
gravitationally bound enviroment (assuming virial equilibrium, and that the
emission arises in a region of $\sim350$-AU radius). This value is about one
order of magnitude larger than the estimated mass of HW2, which is expected to
become a B0.5 star once in ZAMS (Rodriguez et al. \citen{rodriguez1994}).
Fig.~\ref{fig:sio_spectrum} shows a comparison between SiO(2-1) and
\CHTHCN($12_4-11_4$).

We carried out a two-dimensional Gaussian fit of the SiO(2-1) spatial
distribution, for every spectral channel in the velocity range $-25 <$~\vlsr\
$-3.5$~\kms. This corresponds to the spectral interval in which the
signal-to-noise (S/N) ratio of the SiO transition is $\ge 9\sigma$. For
smaller S/N ratios, in fact, the error on the fitted centroid position easily
exceeds 50\%. The result is a distribution of the centroids of SiO emission as
a function of velocity.  Fig.~\ref{fig:ch3cn_nosio}, right panel, shows that
the centroid positions are located in a well-defined two-lobed area, centered
about $0\farcs4$ eastwards of HW2 and of the dust continuum emission
peak. Although the error on every single centroid position is still relatively
large (up to 30\%), as a whole their distribution describes a very clear
velocity trend, with all the emission at \vlsr$< -10$~\kms\ clustering in the
left lobe, and all the emission at \vlsr$> -10$~\kms\ clustering in the right
lobe. This result suggests that \emph{a second molecular outflow is being
ejected in the HW2 region.} Our interpretation is supported by the recent
discovery of an intermediate-mass protostar, surrounded by the hot molecular
core HC (MP05) located in the region between the blue and red lobes of SiO
emission (white triangle in Fig.~\ref{fig:ch3cn_nosio}, right panel), hence a
very likely candidate to be its powering engine. With the current dataset it
is not possible to establish the exact inclination angle of the flow, but the
large velocity spread observed in the SiO(2-1) line, together with the
relatively concentrated spatial distribution of the SiO emission, suggests
that the inclination angle must be high, i.e., that the SiO flow is being
ejected at a small angle with respect to the line of sight.

\subsection{CH$_3$CN}\label{sec:ch3cn}

The dense molecular gas, as traced by CH$_3$CN(12$_3$-11$_3$)
 (Fig.~\ref{fig:ch3cn_nosio}, left panel) appears to be distributed around the
 HW2 position, and elongated in a direction roughly perpendicular to the
 projected direction of the large-scale outflow on the plane of the sky. From
 a morphological point of view, therefore, the data are very suggestive of the
 presence of a $\sim350$-AU-radius disk-like structure around HW2.

\begin{figure}[htbp]
\centering
\includegraphics[bb= 50 250 545 750,clip,angle=-90,width=8cm]{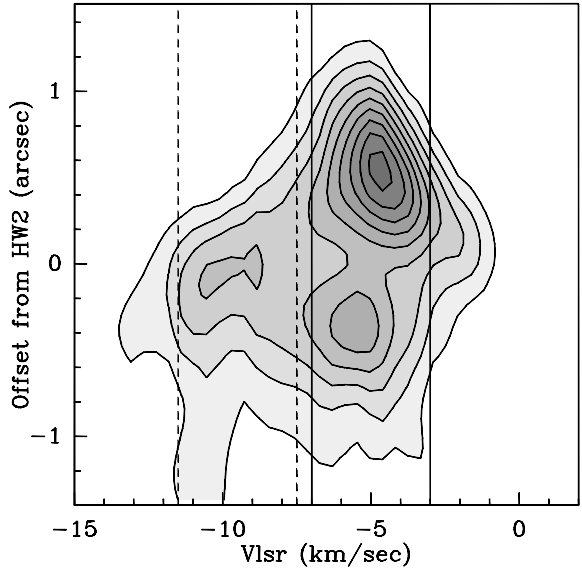}
\caption{Position-velocity plot for \CHTHCN($12_3-11_3$), along the major axis
of the elongated structure (dashed line in Fig.~\ref{fig:ch3cn_nosio}, left
and center panels). Levels range from $3\sigma$ to $27\sigma$ in 3-$\sigma$
steps. The bracketed velocity ranges show the intervals making up the $-5$
(solid) and $-10$ (dashed) \kms\ components, whose spatial distribution is
shown in Fig.~\ref{fig:ch3cn_nosio}, center panel.}\label{fig:ch3cn_pv}
\end{figure}

 On the other hand, as already pointed out by Torrelles et
  al. (\citen{torrelles1999}) based on VLA observations of NH$_3$, the
  kinematical picture describing the dense gas distribution in the region is
  quite complex. A position-velocity cut along the major axis of the elongated
  structure (indicated in Fig.~\ref{fig:ch3cn_nosio}, left and center panels,
  with a dashed line) reveals a velocity spread of about 6~\kms\ (see
  Fig.~\ref{fig:ch3cn_pv}), also observed by Patel et
  al. (\citen{patel2005}). However, the two intensity peaks along the axis
  share roughly the same systemic velocity ($\sim -5$~\kms). The weaker,
  blue-shifted component of emission ($\sim -10$~\kms), appears to trace
  rather the outskirt of a physically separated component than a
  rotation-induced velocity gradient along the axis of the alleged
  ``disk''. The peak of the -10~\kms\ CH$_3$CN emission is spatially and
  kinematically close to the center of the small-scale SiO outflow (see
  Fig.~\ref{fig:ch3cn_nosio}, right panel), it is likely associated to it
  and/or to its exciting source.

\begin{figure}[htbp]
\centering
\subfigure{
\includegraphics[bb= 170 17 520 812,clip,angle=-90,width=8cm]{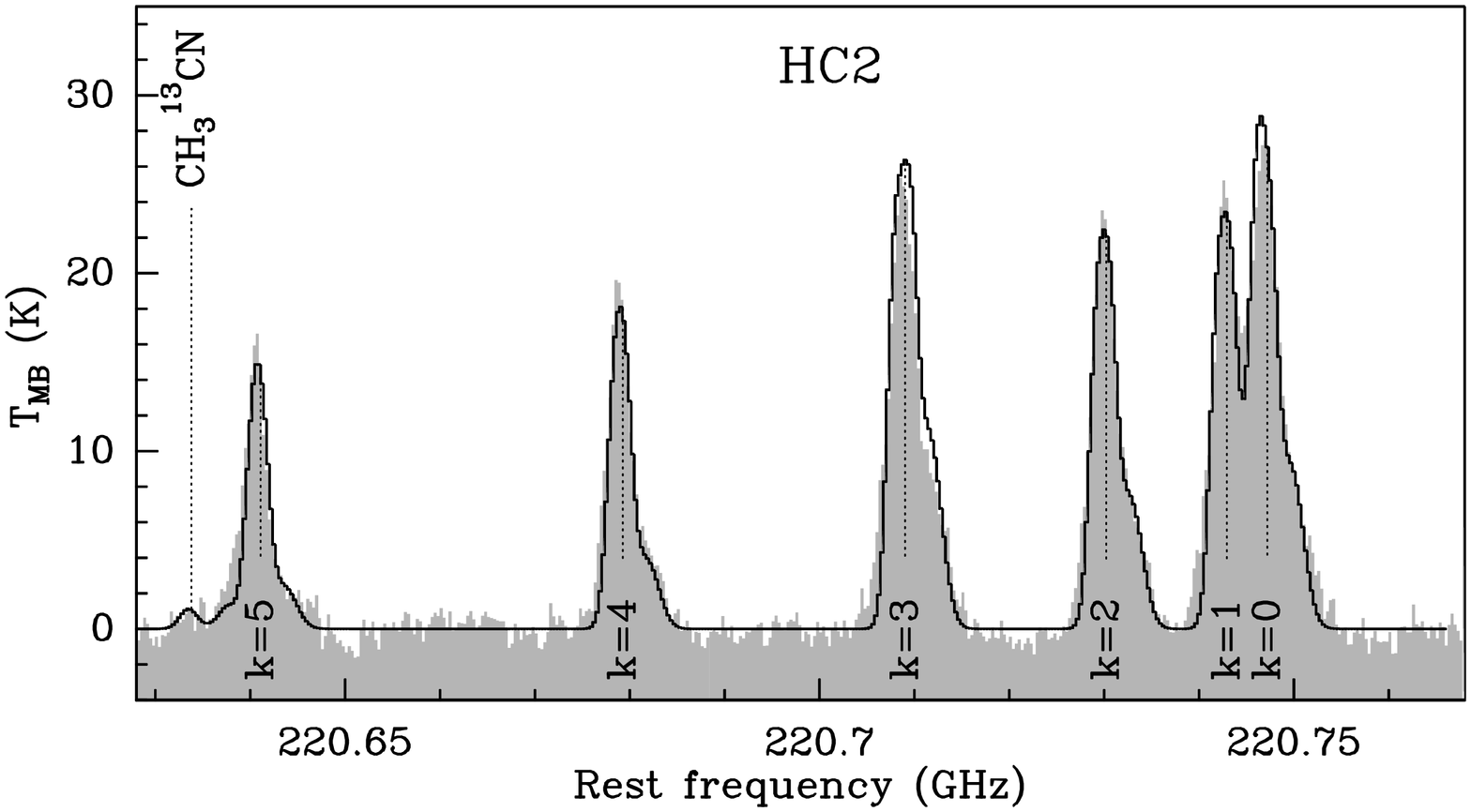}}
\subfigure{
\includegraphics[bb= 170 17 575 812,clip,angle=-90,width=8cm]{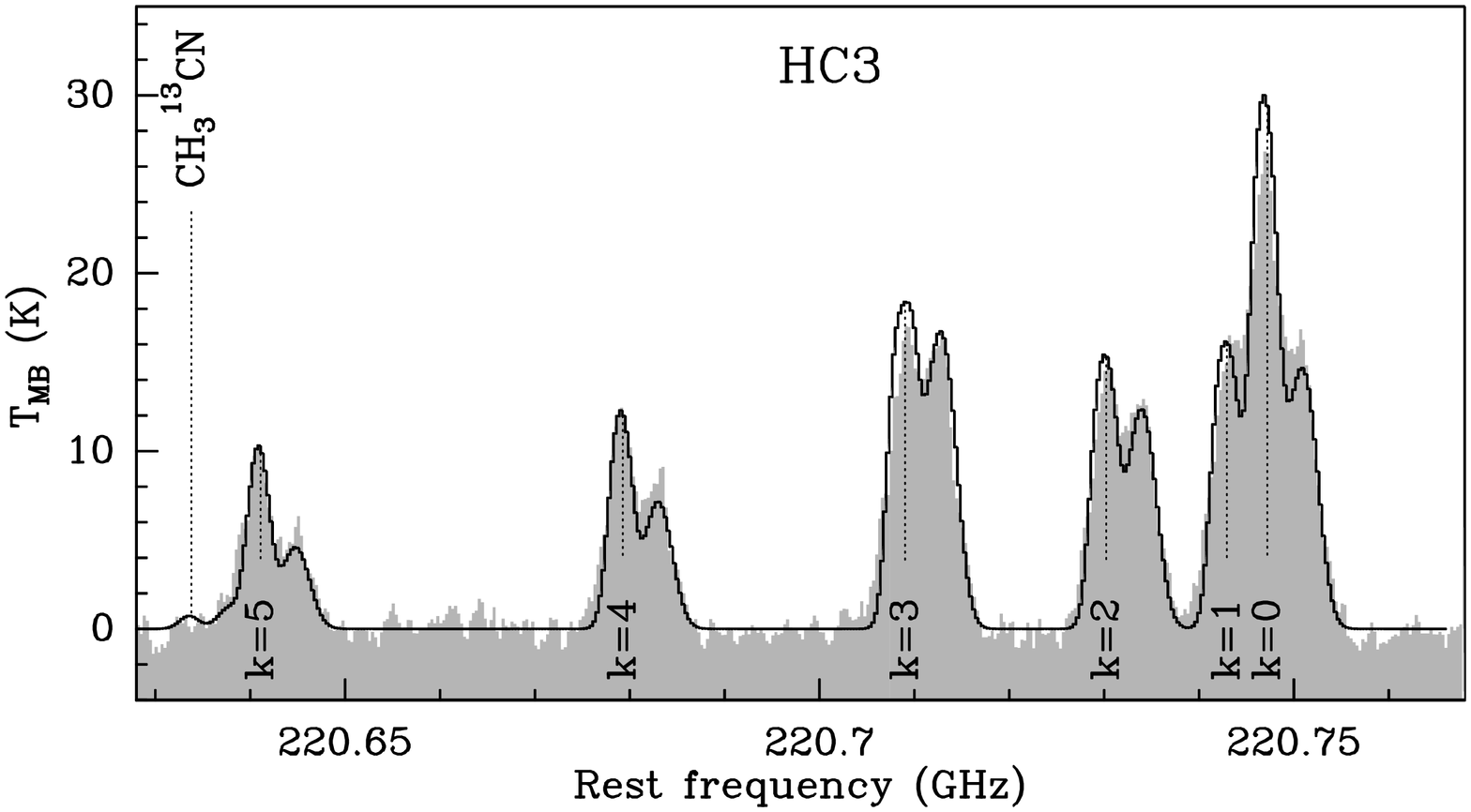}}
\caption{In grey, high-resolution ($\Delta {\rm v}=0.3$~\kms) spectra of the
\CHTHCN(12-11) emission at 200~GHz, towards the HC2 and HC3 positions (see
Fig.~\ref{fig:ch3cn_nosio}, center panel). Overlayed in black are the model
spectra, resulting from the parameters listed in Tab.~\ref{tab:fit}, assuming
LTE approximation and that [$^{12}$CO]$/$[$^{13}$CO]$=60$.}\label{fig:ch3cn_spectra}
\end{figure}

The \CHTHCN\ integrated intensity is dominated by the two $-5$-\kms\ peaks,
which lie respectively about $0\farcs6$ to the northwest, and $0\farcs5$ to
the southeast of the HW2 position. The fact that the two \CHTHCN\ peaks share
roughly the same systemic velocity, is not compatible with the ``rotating
disk'' hypothesis. In what follows, we will treat them as independent
condensations, and to be consistent with the nomenclature introduced by MP05,
we will refer to them respectively as HC2 and
HC3. Fig.~\ref{fig:ch3cn_spectra} compares the spectra observed towards the
two positions. It is clear that, in both cases, both the $-5$- and $-10$-\kms\
components are present along the line of sight, although the contribution from
the latter is more substantial towards HC3, i.e., close to the peak of the
$-10$-\kms\ SiO emission.

We have assumed the LTE approximation to fit the physical parameters
associated with the two different velocity components. All transitions in the
spectrum are fitted simultaneously, in order to take line blending and optical
depth effects properly into account (a detailed description of the method can
be found in Comito et al. \citen{comito2005}). For the $\sim -5$~\kms
component, our fit reveals that the k$=0$ through k$=4$ transitions are
optically thick towards both positions. The data at this velocity can only be
reproduced by including a very compact, hot, dense object in the model. The
emission centered at $\sim -10$-\kms\ can be modeled with a cooler, more
extended component. The results of the fit, for the two positions, are
summarized in Tab.~\ref{tab:fit}.

Although the presence of secondary minima in the $\chi^2$ space is unavoidable
when so many parameters are varied to achieve minimization, in this case the
simultaneous fitting of intensity ratios between optically thick and optically
thin lines, between ortho- and para-\CHTHCN\ transitions, and between $^{12}$C
and $^{13}$C isotopologues of methyl cyanide (see
Fig.~\ref{fig:ch3cn_spectra}), places very stringent constraints on the viable
parameter space, at least as far as the compact ($\sim -5$~\kms) component is
concerned. Note that the rotational temperatures derived in this fashion are
significantly higher than those derived by Patel et
al. (\citen{patel2005}). This discrepancy can be explained with the optical
depth correction in our fit.

\begin{table*}[htbp]
\begin{center}
\begin{tabular}{lccccc}
\hline
\hline
              &    \vlsr   & Source size &  N(\CHTHCN)              &  T$_{\rm
  rot}$ & $\Delta {\rm v}$  \\

              &     (\kms)   &            & (\cmsq)                  &   (K)         &  (\kms)    \\
\hline
HC2    &    -4.5      & $0\farcs3$ & $\sim3\times10^{16}$  &   250        &  2.9     \\
           &    -8.9      & $1\arcsec$ & $\sim 8\times10^{14}$ &   150        &  4.0    \\   
HC3    &    -4.2      & $0\farcs25$& $\sim3\times10^{16}$  &   250        &  3.2    \\
           &    -10.0     & $0\farcs45$& $\sim5\times10^{15}$  &   150        &  4.5   \\
\hline 
\end{tabular}
\caption{LTE model results for the \CHTHCN\ emission towards the HC2 and HC3
cores (Fig.~\ref{fig:ch3cn_spectra}). For a discussion on the error estimate,
cf. Comito et al. \citen{comito2005}.}\label{tab:fit}
\end{center}
\end{table*}

\section{Discussion}\label{sec:discussion}

\begin{figure}[htbp]
\centering
\includegraphics[bb= 20 220 570 810,clip,angle=-90,width=9.5cm]{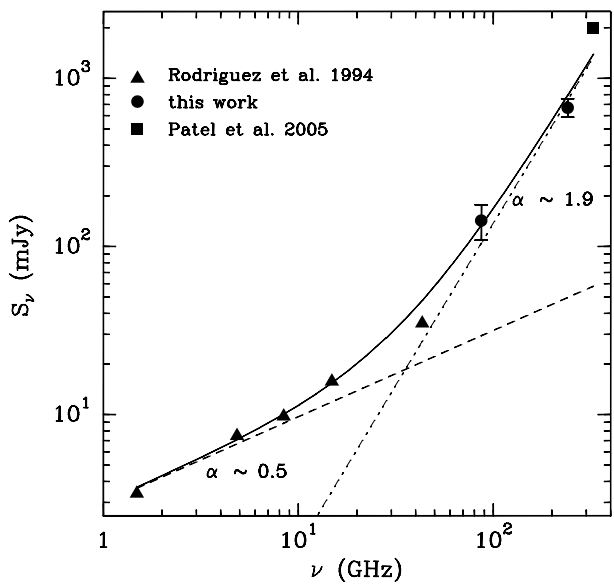}
\caption{ Flux density as a function of frequency for Cep-A HW2. For the
Rodr\'{\i}guez et al. (\citen{rodriguez1994}, triangles) and Patel et
al. (\citen{patel2005}, square) data points, the errorbars fall within the
symbols and are therefore not visible. The solid curve shows the 2-component
least-squares fit of the data, as descrived in the text. The single components
resulting from the fit are also plotted separately (dashed and dashed-dotted
lines).}\label{fig:continuum}
\end{figure}

The observed elongation of the molecular gas distribution around HW2, over a
radius of $\sim0\farcs5$ ($\sim 360$~AU), appears to be due to the projected
superposition, on the plane of the sky, of at least three protostellar
objects, of which at least one is triggering a molecular outflow at a small
angle with respect to the line of sight (\S~\ref{sec:sio},
\S~\ref{sec:ch3cn}). All lines in our dataset are consistent with this
interpretation. The distribution of molecular gas around HW2 can, on a
$1\arcsec$ scale, be interpreted as a cluster of high- and intermediate-mass
protostars in the Cepheus A HW2 region.
The analysis of the \CHTHCN\ spectra (\S~\ref{sec:ch3cn}) suggests the
presence of internally heated compact hot-core-type objects like HC, likely
hosting protostellar objects, although the 1-mm continuum emission peaks on
the HW2 position and does not show any secondary clumps. This may be due to
insufficient dynamic range in our data, if the contribution, to the 241-GHz
continuum, of free-free emission from the HW2 thermal jet is
large. Fig.~\ref{fig:continuum} shows the variation of the measured HW2
continuum flux density as a function of frequency, $S_{\nu}$, between 1.5 and
327 GHz (data points from: Rodr\'{\i}guez et al. \citen{rodriguez1994}; this
work; Patel et al. \citen{patel2005}). A two-component least-squares fit of
the data yields $S_{\rm jet} \propto \nu^{(0.51\pm0.12)}$ (consistent with the
value inferred by Rodr\'{\i}guez et al. \citen{rodriguez1994}, and with the
theoretical predictions for the radio continuum spectrum of a confined thermal
jet, Reynolds \citen{reynolds1986}) and $S_{\rm (sub)mm} \propto
\nu^{(1.92\pm0.12)}$ (dashed and dashed-dotted lines respectively in
Fig.~\ref{fig:continuum}), where $S_{\rm jet}+ S_{\rm (sub)mm} = S_{\nu}$
(solid curve in Fig.~\ref{fig:continuum}). Based on this estimate, the thermal
jet (free-free) contribution at 241 GHz should be $\sim 10\%$ of the total
flux.

However, the flux density variation in the (sub)mm-wavelength portion of the
spectrum increases basically on a Rayleigh-Jeans slope, suggesting the
presence of optically thick emission from an unresolved (with our best spatial
resolution, at 241 GHz with PdBI) continuum source, whose size thus cannot be
larger than $\sim 0\farcs6$. Although we cannot determine the nature of this
compact source, it makes sense to hypothesize that it can be described by {\it
i)}, dust emission, and/or {\it ii)}, free-free emission from a \HII\ region
(for example associated to a photoevaporating disk). For case {\it i)}, we
adopt Beckwith et al.'s (\citen{beckwith1990}) values for the mass absorption
coefficient, $\kappa_{\nu} = 0.1(\nu/10^{12} {\rm Hz})^{\beta}$, with $\beta =
1$, to estimate the lower mass limit for an object with size $\theta_{\rm
dust} = 0\farcs5$ ($\sim 363$~AU) to produce optically thick emission at
87~GHz: $M_{\tau_{\rm dust} = 1} \geq 1$~\Msun. Based on the peak flux at 87,
241 (this work) and 327 GHz (Patel et al. \citen{patel2005}), the above value
of $\theta_{\rm dust}$ yields a brightness temperature, $T_{\rm B} \simeq
80$~K. Smaller source sizes would lead to higher intrinsic temperatures and
lower mass limits.  In case {\it ii)}, we assume $T_{\rm B} = 10^4$~K (typical
for \HII\ regions), which would translate into a source size of $\theta_{\rm
free-free} \simeq 0\farcs04$ ($\sim 30$~AU).  The continuum source VLA-mm
(Curiel et al. \citen{curiel2002}), which is located $\sim 0\farcs15$ south of
HW2 and whose size cannot be larger than $30$~AU, displays a much too weak
emission at cm and mm wavelengths, and we estimate its contribution to cover
for at most 5\% of the observed (sub)mm flux.

In other words, we cannot discard any of the two hypotheses for the observed
optically thick continuum emission. At higher frequencies, the spectral index
would get flatter if the optically thick emission were only due to free-free
emission, while it would remain the same for dust. Observations with a
resolution of $<0.5\arcsec$, which are within the reach of present day
interferometers, would be able to shed more light on the nature of this
object.

As these estimates show, our conclusions do not rule out at all the existence,
on a smaller scale, of an accretion disk around HW2, which in fact is to be
expected, based on the very presence of the HW2 jet. In fact, recent 7-mm VLA
observations of SO$_2$ have led Jim{\'e}nez-Serra et
al. (\citen{jimenezserra2007}) to claim the detection of a disk-like structure
with a size of $600 \times 100$~AU, roughly centered on the HW2 position, part
of which may be photoevaporating. Although spatially almost coexistent on the
plane of the sky, this structure is characterized by a different \vlsr\
($-7.3$~\kms, as opposed to $\sim -5$~\kms) and apparently a different
chemistry from the molecular gas traced by \CHTHCN. Our above estimates on the
nature of the black-body emission in the (sub)mm regime are all consistent
with Jim{\'e}nez-Serra et al.'s conclusions.

Overall, the Cepheus A HW2 allows, due to its proximity, a view into the heart
of a massive star forming region. The emerging picture is anything but simple:
including the sources detected by Curiel et al. (\citen{curiel2002}), and the
hot cores HC, HC2 and HC3, at least 6 probable young stellar or protostellar
objects are located within a radius of $1\arcsec$ or 725~AU. It remains an
open issue whether, under such circumstances we can expect to observe a
classical accretion disk feeding a single central star, or rather some kind of
circum-cluster disk or ring-like structure (analogous perhaps to circumbinary
rings like the one around GG Tau, Guilloteau et al. \citen{guilloteau1999}),
and what such a structure may look like, both from a morphological and from a
kinematical point of view.  Higher spatial resolution is needed, but the
challenge is to identify the right chemical tracer to investigate the
structures one is interested in.  \CHTHCN, otherwise considered a reasonably
good disk tracer (e.g. for the disk in IRAS 20126, Cesaroni et
al. \citen{cesa1997}), does not seem to trace the disk-like structure seen by
Jim{\'e}nez-Serra et al. (\citen{jimenezserra2007}) at all.

Another issue is the physical location of the $-10$~\kms\ molecular
component. Though it seems likely that the peak of $-10$-\kms\ \CHTHCN\
emission is associated with the powering source of the small-scale SiO
outflow, its connection to the somewhat more extended molecular emission at
this systemic velocity (cf. Brogan et al. \citen{brogan2007}) remains to be
confirmed.


\begin{acknowledgements}
The authors are grateful to the IRAM staff in Grenoble, particularly to
H.~Wiesemeyer, J.~M. Winters and R.~Neri, for their support in the data
calibration process. An anonymous referee has given a significant contribution
to the improvement of this paper. CC and PS have enjoyed many fruitful
discussions with Malcolm Walmsley. JMP and IJS acknowledge the support
provided through projects number ESP2004-00665 and S-0505$/$ESP-0277
(ASTROCAM).
\end{acknowledgements}

\end{document}